\documentclass[fleqn,10pt]{wlscirep}
\usepackage[utf8]{inputenc}
\usepackage[T1]{fontenc}
\usepackage{bm}
\usepackage{float}
\usepackage{multirow}
\title{A Data-Driven Regional Model for Skillful Medium Range Typhoon Prediction}

\author[1,2]{Zeyi Niu}
\author[1,*]{Wei Huang}
\author[1]{Sirong Huang}
\author[3]{Zhuo Wang}
\author[2]{Mu Mu}
\author[1]{Mengqi Yang}
\author[1]{Xinhai Han}
\author[1]{Haofei Sun}
\author[1]{Zhaoyang Huo}
\author[2]{Bo Qin}

\affil[1]{Shanghai Typhoon Institute, and Key Laboratory of Numerical Modeling for Tropical Cyclone of the China Meteorological Administration, Shanghai, China}

\affil[2]{Department of Atmospheric and Oceanic Sciences and Institute of Atmospheric Sciences, Fudan University, Shanghai 200438, China}

\affil[3]{Department of Atmospheric Sciences, University of Illinois at Urbana--Champaign, Urbana, Illinois, USA}
\affil[*]{e-mail: huangw@typhoon.org.cn}

\begin{abstract}
Accurate prediction of tropical cyclones remains a major challenge for both numerical weather prediction and emerging artificial intelligence weather prediction (AIWP) systems. While recent global AI models have demonstrated strong skill in large-scale circulation prediction, they often struggle to represent the mesoscale structures critical for tropical cyclone intensity and precipitation. Here we develop the Hybrid Intelligent Typhoon System (HITS), a regional AI forecasting framework for medium-range typhoon prediction over the Asia–Pacific region, trained on a newly constructed 9-km high-resolution typhoon reanalysis dataset. The model combines regional autoregressive prediction with large-scale dynamical constraints from the state-of-the-art ECMWF Artificial Intelligence Forecasting System (AIFS), allowing it to remain dynamically consistent with the evolving large-scale circulation while resolving mesoscale structures. HITS is further extended with a structure-aware perceptual training strategy (HITS-LPIPS) that improves the representation of convective and typhoon rainband structures. Experiments show that the hybrid framework substantially improves precipitation structure and typhoon intensity forecasts compared with both purely autoregressive regional AI models and standalone AI downscaling approaches. In particular, HITS-LPIPS reduces intensity errors by up to 47.8\% relative to AIFS at a 72-hour lead time and produces a near-unbiased wind–pressure relationship for simulated typhoons. These results demonstrate that dynamically constrained regional AI systems provide a promising pathway for improving medium-range typhoon prediction. 
\end{abstract}
\begin{document}

\flushbottom
\maketitle

\thispagestyle{empty}

\section*{Introduction}
Global medium-range numerical weather prediction (NWP) is a core component of the modern forecasting system\textsuperscript{1,2}. Many high-impact weather events, including tropical cyclones, heat waves and heavy precipitation leading to flooding, occur on this timescale\textsuperscript{3,4}. Global NWP models not only provide operational forecasts but also supply boundary conditions for regional models, generate (re)analysis datasets, and serve as key tools for understanding atmospheric processes\textsuperscript{5,6}. Modern NWP systems are built on discretizations of the governing equations of fluid dynamics and thermodynamics\textsuperscript{7}. At present, most global models operate at kilometre-scale horizontal resolution\textsuperscript{8}; for example, the high-resolution system (HRES) of the European Centre for Medium-Range Weather Forecasts (ECMWF) runs at approximately $0.1^\circ$ ($\sim$9 km). Nevertheless, many key physical processes, such as cloud microphysics and radiation, remain unresolved and must be represented using parameterization schemes\textsuperscript{9,10}. Another essential component of NWP is data assimilation, which combines model forecasts with observations to produce the best estimate of the atmospheric state\textsuperscript{11--13}. Over the past decades, global NWP skill has steadily improved owing to advances in computing power, observing systems, data assimilation techniques and physical parameterizations\textsuperscript{14}. Despite these advances, the theoretical predictability limit of mid-latitude weather is estimated to be about 15 days, whereas the current practical limit is around 10 days; roughly half of the remaining potential predictability is associated with model improvements and the other half with more accurate initial conditions\textsuperscript{15,16}. Meanwhile, the development of traditional NWP is increasingly constrained by slowing growth in computational resources and the rising complexity of physical models. Operational centers often face a trade-off between increasing model resolution and enlarging ensemble size, which limits further progress\textsuperscript{17}.

To address these challenges, researchers have begun exploring alternative forecasting paradigms, particularly the use of artificial intelligence (AI) for weather prediction\textsuperscript{18--21}. Early attempts were limited by computational resources and algorithmic capability and did not outperform ECMWF-HRES. This situation changed in 2023 with the release of the Transformer-based Pangu-Weather model\textsuperscript{22}, which for the first time achieved higher forecast skill than ECMWF-HRES while reducing computational cost by more than four orders of magnitude. Since then, several AI Weather Prediction (AIWP) models have been developed, including FourCastNet\textsuperscript{23}, GraphCast\textsuperscript{24}, FengWu\textsuperscript{25}, FuXi\textsuperscript{26} and the ECMWF Artificial Intelligence Forecasting System (AIFS)\textsuperscript{27}, marking the beginning of a new paradigm in weather prediction\textsuperscript{28--31}. Despite their strong performance in tropical cyclone track forecasts, current AIWP models still show notable limitations in predicting extreme events, particularly tropical cyclone intensity, which is often systematically underestimated. A key reason is that most AIWP models are trained on the ERA5 reanalysis dataset with a horizontal resolution of approximately 25 km, which cannot resolve the mesoscale structures critical for tropical cyclone intensity development\textsuperscript{32,33}.

To improve typhoon intensity prediction, several studies have coupled AIWP forecasts with NWP models to form AI--physics hybrid systems through dynamical downscaling\textsuperscript{34--38}. Although this approach can enhance typhoon intensity forecasts, it increases computational cost and reduces operational efficiency. Alternatively, some studies perform AI-based downscaling to map global AIWP forecasts to higher-resolution fields. For example, the CorrDiff model downscales ERA5 to 2 km resolution over Taiwan\textsuperscript{39}, while other studies project AIWP forecasts onto satellite-derived wind fields such as Synthetic Aperture Radar observations\textsuperscript{40}. In parallel, regional AIWP models based on high-resolution reanalysis datasets have also been explored, including hierarchical graph neural network approaches and boundary-condition-aware regional AI models\textsuperscript{41--45}. In addition, NVIDIA's StormCast represents an early attempt at AI-based convection-scale forecasting\textsuperscript{46}. StormCast adopts a conditional diffusion framework with a U-Net denoising network and large-scale forcing from NCEP Global Forecast System (GFS) to produce convection-scale forecasts at $\sim$3 km resolution. An upgraded system, HRRRCast\textsuperscript{47}, further improves generalization and reduces autoregressive error accumulation through full-domain training, HRRR analysis targets, and multi-lead-time training, achieving higher precipitation skill than the physical HRRR model.

However, most existing regional AIWP models focus on short-range weather prediction with lead times typically shorter than 48 hours, which cannot meet the requirements of medium-range tropical cyclone forecasts. To address this gap, this study proposes a regional AI forecasting model, the Hybrid Intelligent Typhoon System (HITS), designed for five-day typhoon prediction over the Asia--Pacific region. This study makes three main contributions through the development of HITS, which has three key characteristics that distinguish it from existing regional AI models.

\begin{enumerate}
\item \textbf{Regional typhoon-focused forecasting.} HITS covers the entire Asia--Pacific region and is designed as a unified regional AIWP model for typhoons, producing forecasts from 0 to 120 hours. The model is trained on a high-resolution typhoon reanalysis dataset (HiRes)\textsuperscript{48}, enabling it to better learn typhoon structure and intensity evolution over the western North Pacific. Notably, compared with ERA5, HiRes significantly improves the representation of typhoon intensity and wind structure. In contrast, existing regional AIWP models such as StormCast mainly target short-range precipitation forecasting (0--48 hours) with a limited area.

\item \textbf{Hybrid forecasting--downscaling framework.} HITS integrates forecasting and downscaling within a unified architecture. Unlike StormCast, the large-scale forecast at the future lead time ($t+6$ h) is introduced into the network via cross-attention as a dynamical constraint. This design helps maintain consistency with the evolving large-scale circulation during long-range autoregressive predictions. Our results show that this hybrid framework outperforms both purely autoregressive AI forecasting and standalone AI downscaling approaches in predicting precipitation and typhoon intensity.

\item \textbf{Structure-aware training strategy.} To preserve the mesoscale physical structures of typhoon rather than generating random small-scale details, the model is trained with a structure-aware perceptual loss instead of the commonly used diffusion-based approach. Diffusion may produce random small-scale noise
LPIPS preserves physically consistent structures. The resulting HITS-LPIPS model shows improved representation of key structural features such as spiral rainbands.
\end{enumerate}

\section*{Results}

\subsection*{Experiments}
To demonstrate the advantages of the hybrid AI framework and the effectiveness of multi-scale loss functions, four model configurations are designed: ISTM, CTL, HITS, and HITS-LPIPS.ISTM is a two-stage downscaling framework that refines AIFS forecasts using a diffusion-based generative module, focusing on surface variables to enhance fine-scale spatial structures while maintaining large-scale consistency from AIFS. CTL serves as the baseline regional AI model and adopts a U-Transformer autoregressive architecture trained solely on the HiRes dataset without external boundary. HITS extends CTL as a hybrid forecasting--downscaling model by incorporating large-scale fields from AIFS forecasts through a cross-attention mechanism, enabling the model to perform regional prediction while maintaining dynamical consistency with the global-scale atmospheric circulation. HITS-LPIPS further improves HITS by introducing a multi-scale perceptual constraint during training, where a Learned Perceptual Image Patch Similarity (LPIPS) term is added to the reconstruction loss to better represent mesoscale and convective-scale structures and improve the realism of typhoon intensity and precipitation-related features.It should be noted that all AI models in this study, including AIFS, are evaluated using the same sample set. All designed models are initialized at 0000 and 1200~UTC each day and produce forecasts up to 120~hours.Besides, our focus is regional AI framework comparison rather than global model benchmarking.
\subsection*{Overall forecast performance}
The forecast performance of CTL, HITS, HITS-LPIPS, ISTM, and ECMWF-AIFS is compared in terms of RMSE for multiple variables over lead times from 6 to 120~hours (Fig.~1). The evaluated variables include temperature ($t$), zonal and meridional winds ($u$ and $v$), and specific humidity ($q_v$) at four pressure levels (250~hPa, 500~hPa, 850~hPa, and 925~hPa), as well as four surface variables: 2-m temperature ($t_2$), 10-m zonal wind ($u_{10}$), 10-m meridional wind ($v_{10}$), and mean sea level pressure (MSLP).

Overall, the RMSE of CTL (black) increases most rapidly with forecast lead time and becomes substantially larger at 120~hours. This result indicates that the regional AI autoregressive model without external boundary conditions suffers from pronounced error accumulation in medium-range forecasts. In contrast, HITS (blue) shows markedly lower RMSE than CTL, indicating that introducing large-scale constraints from AIFS effectively anchors the evolution of the synoptic-scale circulation within the regional model domain. After incorporating the structure-aware LPIPS loss, HITS-LPIPS (red) further reduces RMSE relative to HITS for nearly all variables and lead times.

ISTM (green) is a downscaling AI model designed for surface variables. Its overall forecast performance is better than AIFS but remains slightly inferior to HITS and HITS-LPIPS (Fig.~1e1--e4). A comparison between HITS-LPIPS and AIFS further shows that AIFS performs noticeably better for variables at 250~hPa. This advantage likely arises from the stronger large-scale prediction capability of AIFS in the upper troposphere, whereas HITS-LPIPS only incorporates AIFS information at 850~hPa and 500~hPa, resulting in relatively weaker performance at 250~hPa.

However, at 500~hPa, 850~hPa, 925~hPa, and for surface variables, AIFS generally exhibits larger RMSE than HITS-LPIPS (e.g., t925, v925, t2, and MSLP). These results indicate that training and evaluation based on high-resolution regional reanalysis significantly improve the prediction of near-surface atmospheric variables. In addition, CTL, HITS, and HITS-LPIPS, which are driven by high-resolution initial conditions, outperform AIFS and ISTM during the first $\sim$18~hours of forecast lead time, particularly for surface variables. This result highlights the importance of high-resolution initial fields for short-range autoregressive prediction.

Typhoon track prediction largely depends on the accuracy of the 500~hPa circulation field. As shown in Fig.~1, HITS-LPIPS and AIFS exhibit comparable forecast skill at 500~hPa. Furthermore, the spatial distribution of 500~hPa geopotential height ($z_{500}$) shown in Fig.~S1 indicates that the forecasts from HITS-LPIPS are generally similar to those from AIFS, while CTL shows noticeably larger deviations from the reference fields.
\begin{figure}[!htbp]
\centering
\includegraphics[width=\linewidth]{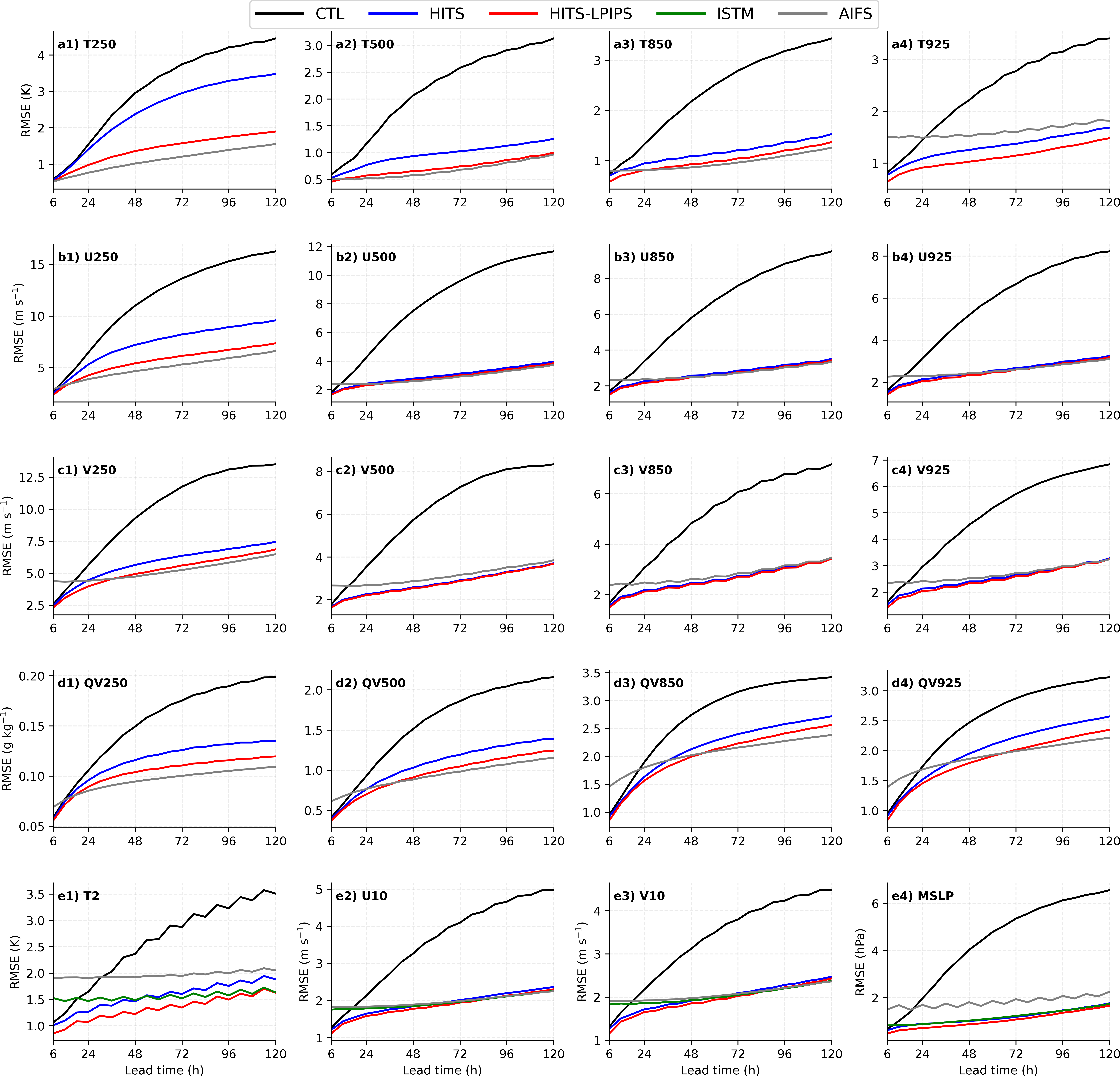}
\caption{\textbf{Lead-time dependence of forecast skill for atmospheric variables across different experiments.} Panels show verification against HiRes as a function of lead time from 6 to 120~h, evaluated over the testing period from June to November 2025. Rows (a1--a4), (b1--b4), (c1--c4), and (d1--d4) show the root mean square error (RMSE) of temperature, zonal wind, meridional wind, and specific humidity at 250, 500, 850, and 925~hPa, respectively. Specific humidity is given in g~kg$^{-1}$. The last row (e1--e4) shows surface variables, including 2~m temperature, 10~m zonal wind, 10~m meridional wind, and mean sea level pressure (MSLP). Black, blue, red, green, and gray curves denote CTL, HITS, HITS-LPIPS, ISTM, and AIFS, respectively.}
\label{fig:figure1}
\end{figure}

\subsection*{Precipitation forecasts (radar)}
The precipitation forecast performance of different models is further evaluated. In this study, precipitation is represented by composite radar reflectivity, consistent with the evaluation approach used in the StormCast model. Figure~2 presents the forecast results and skill verification for a representative summer convective precipitation event in China initialized at 0000~UTC on 11~August~2025. Figures~2a--e compare the spatial distributions of composite radar reflectivity from different models at forecast lead times of 6--36~hours.

From the spatial distributions, HiRes, used as the ground truth (Fig.~2a), clearly depicts the organized structure of mesoscale convective bands and the embedded intense convective cores. Regions with large reflectivity values exhibit distinct banded and clustered patterns with strong spatial continuity. Overall, CTL, HITS, and HITS-LPIPS successfully reproduce the general location, morphology, and main convective centers of the precipitation system. However, HITS-LPIPS (Fig.~2b) better captures the convective structures, the locations of strong echo cores ($>35$~dBZ), and the clustered distribution characteristics compared with HITS and CTL. HITS (Fig.~2c) generally predicts the convective precipitation area reasonably well, but the predicted structures appear smoother. CTL (Fig.~2d) maintains the approximate location of the convective band at short lead times, but as the forecast lead time increases, the precipitation region gradually shifts and the strong reflectivity regions become artificially elongated, indicating the accumulation of autoregressive errors. ISTM (Fig.~2e) underestimates both the intensity and spatial extent of convection owing to the lack of high-resolution initial conditions, and its predicted fine-scale structures appear overly diffused and remain inconsistent with the reference fields.

A similar analysis is conducted for the extreme rainfall event over Korea on 16~July (Fig.~S2), and the conclusions remain consistent. HITS-LPIPS reproduces both the intensity and structure of the convective system well. In contrast, HITS produces overly smooth structures, CTL shows smoothed features and location shifts at longer lead times, and ISTM significantly underestimates convective intensity.

Furthermore, the average Fractions Skill Score (FSS) is calculated for multiple precipitation events during June--December~2025 to evaluate the evolution of precipitation forecast skill with lead time. Four reflectivity thresholds (10~dBZ, 20~dBZ, 30~dBZ, and 40~dBZ) are used for verification (Fig.~2f). The results show that CTL achieves higher FSS values than ISTM during the first $\sim$42~hours, whereas after 42~hours ISTM performs better than CTL. This behavior further indicates that short-range forecasts are primarily controlled by the initial conditions, whereas longer-range forecasts depend more strongly on the large-scale circulation. By integrating the advantages of both approaches, HITS consistently outperforms CTL and ISTM. Furthermore, the introduction of the structure-aware loss function in HITS-LPIPS further enhances the prediction of convective structures, resulting in the best precipitation forecast performance among all models.

Notably, for the highest threshold ($>40$~dBZ), the forecast skill of HITS-LPIPS is more than twice that of the other models, highlighting its improved capability in predicting extreme precipitation. Overall, the results demonstrate three key findings. First, the introduction of large-scale dynamical constraints in HITS significantly improves the spatial consistency of mesoscale precipitation structures. Second, incorporating the structure-aware loss function in HITS-LPIPS further enhances the representation of intense convective cores. Third, both the purely autoregressive regional model (CTL) and the pure downscaling model (ISTM) show relatively limited performance at longer lead times or higher reflectivity thresholds. These results confirm the clear advantage of the ``dynamical constraint + structural enhancement'' hybrid framework for convective-scale precipitation forecasting and highlight the critical role of structure-aware training objectives in improving high-reflectivity precipitation prediction skill.
\begin{figure}[!htbp]
\centering
\includegraphics[width=\linewidth]{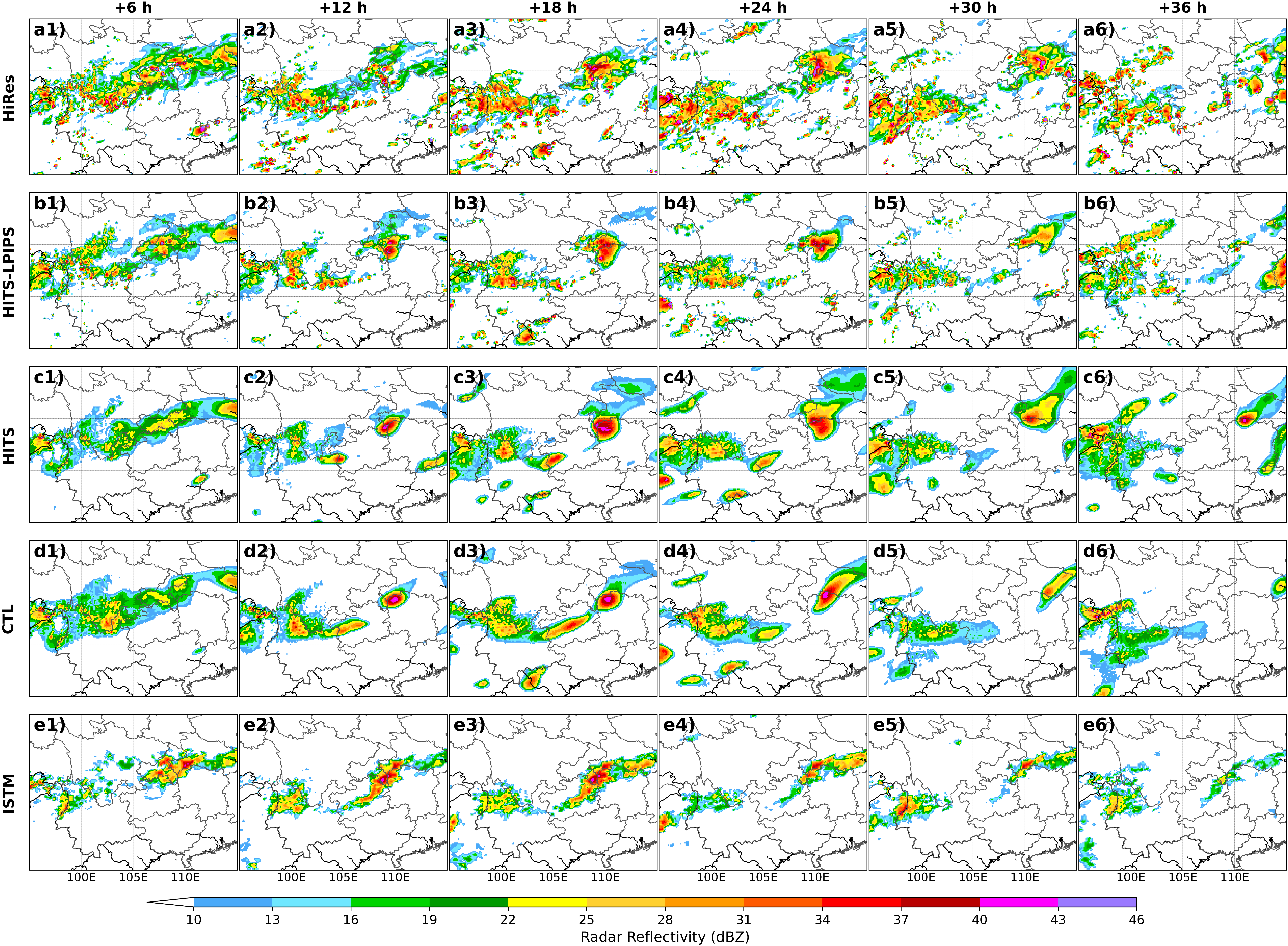}

\vspace{0.5em}

\includegraphics[width=\linewidth]{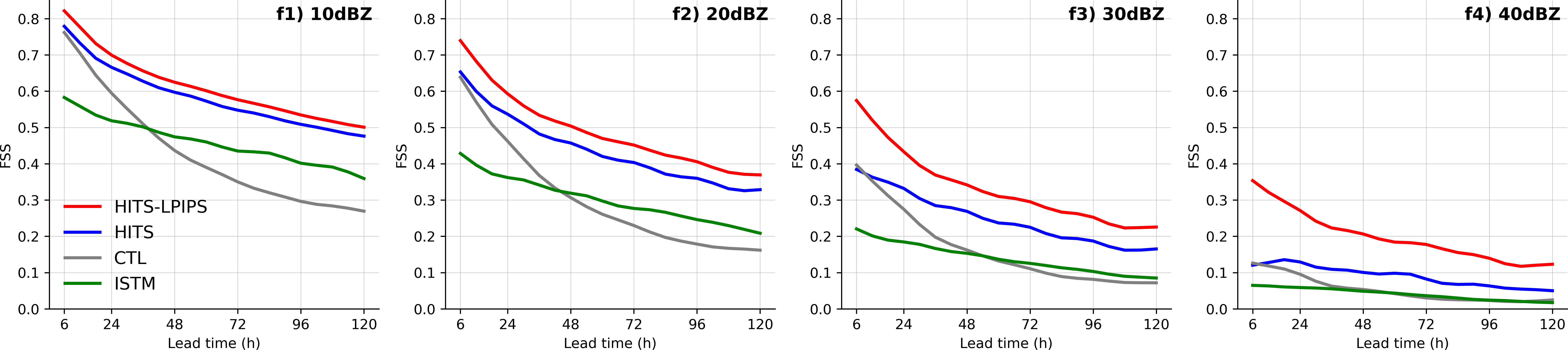}
\caption{\textbf{A typical summer convective precipitation event and evaluation of precipitation forecast skill across different models in 2025.} Columns show forecast lead times of 6~h, 12~h, 18~h, 24~h, 30~h, and 36~h, respectively, initialized at 0000~UTC on 11~August~2025. Rows correspond to composite radar reflectivity forecasts from (a1--a6) HiRes, (b1--b6) HITS-LPIPS, (c1--c6) HITS, (d1--d6) CTL, and (e1--e6) ISTM. Panels (f1--f4) present the fractions skill score (FSS) of composite radar reflectivity as a function of lead time (6--120~h) for thresholds above 10~dBZ, 20~dBZ, 30~dBZ, and 40~dBZ, respectively. The FSS values are averaged over June--December~2025, comparing CTL (grey), HITS (blue), HITS-LPIPS (red), and ISTM (green). Higher FSS indicates improved spatial agreement of reflectivity exceedance patterns.}
\label{fig:figure2}
\end{figure}

\subsection*{Typhoon track, intensity, and structure forecasts}
This subsection evaluates the forecast performance of different models for typhoons. Figure~3 presents the evolution of composite radar reflectivity for Typhoon Danas (2025), initialized at 0000~UTC on 5~July~2025, at different forecast lead times. Overall, HITS-LPIPS provides more accurate predictions of both typhoon position and structural features than the other models across most lead times. In particular, at the 24-hour forecast lead time, HITS-LPIPS (Fig.~3b3) successfully reproduces isolated convective cells embedded within the outer spiral rainbands. In contrast, the rainband structures predicted by HITS (Fig.~3c3) and CTL (Fig.~3d3) appear overly smooth, indicating insufficient representation of small-scale convective structures. 

For ISTM, the composite radar reflectivity is obtained by downscaling the total column water content from AIFS. Because the AIFS total water content fields are themselves relatively smooth, they cannot provide sufficient small-scale information for convective structures. As a result, the rainband details generated by ISTM appear largely random and lack coherent organization. This behavior is also consistent with a known limitation of diffusion-based downscaling models, which can sometimes generate artificial small-scale echoes that are not physically consistent with observations.

A similar comparison is performed for Super Typhoon Ragasa (2025), and HITS-LPIPS again shows better structural prediction of the typhoon system than the other models (Fig.~S3). Further comparisons of track and intensity forecasts for Super Typhoon Ragasa (Fig.~S4) show that, except for CTL, the different experiments produce very similar track forecasts. This result is consistent with the large-scale circulation predicted by AIFS. In terms of intensity, HITS-LPIPS produces the strongest intensity forecasts among the tested models. However, it still underestimates the intensity of the super typhoon by approximately 10--15~m~s$^{-1}$, indicating that the HiRes (9~km) simulations used for training remain relatively weak in representing extreme typhoon intensity.
\begin{figure}[!htbp]
\centering
\includegraphics[width=\linewidth]{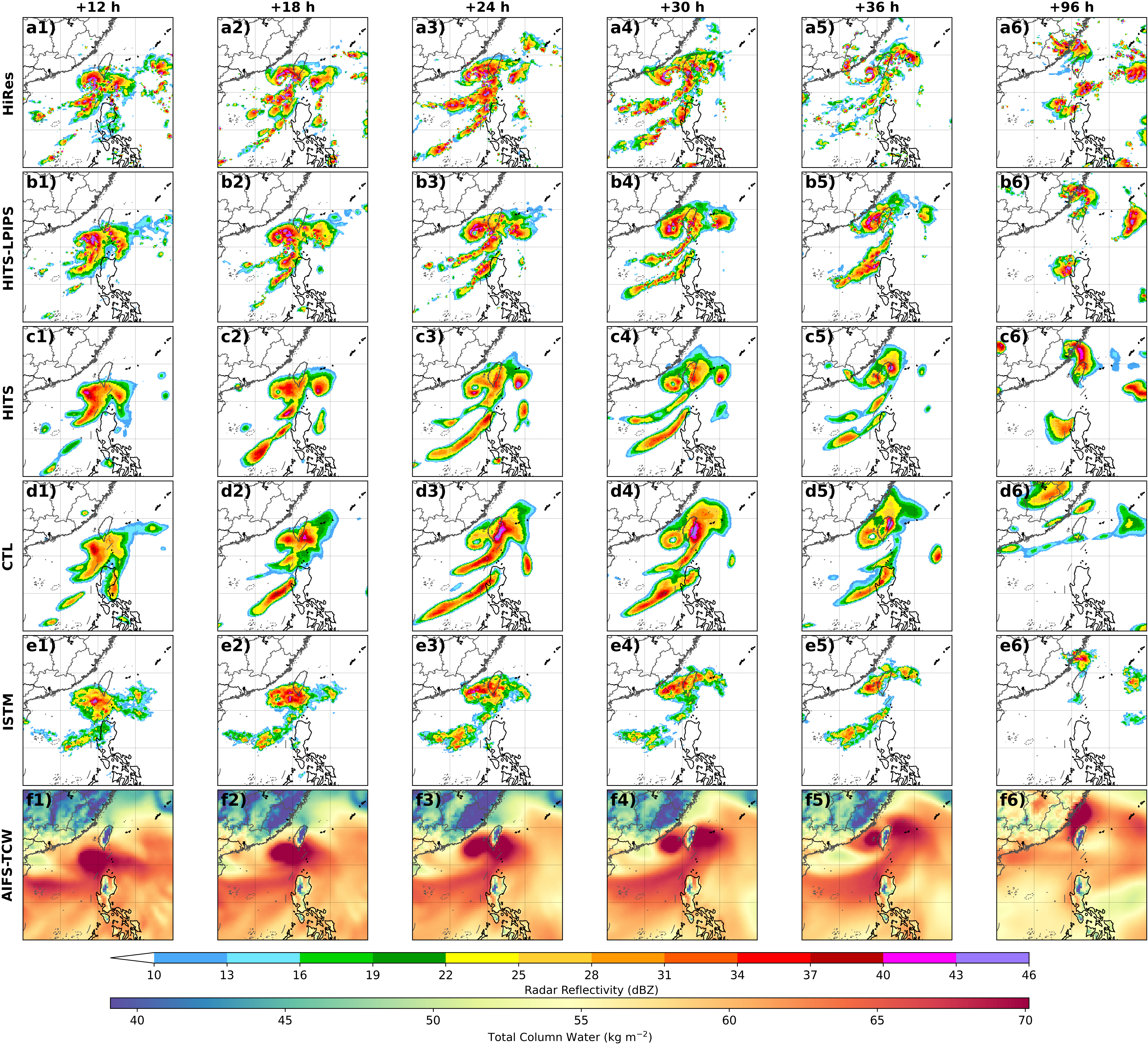}
\caption{\textbf{Comparison of composite radar reflectivity forecasts for Typhoon Danas (2025) across different experiments.} Rows correspond to (a1--a6) HiRes, (b1--b6) HITS-LPIPS, (c1--c6) HITS, (d1--d6) CTL, (e1--e6) ISTM, and (f1--f6) AIFS-TCW. Columns show forecast lead times of 12~h, 18~h, 24~h, 30~h, 36~h, and 96~h from forecasts initialized at 0000~UTC on 5~July~2025.}
\label{fig:figure3}
\end{figure}
To further evaluate overall forecast performance, batch verification of typhoon track and intensity forecasts during the 2025 typhoon season is conducted (Fig.~4). Figure~4a shows the CMA best-track dataset for Typhoons~1--27 over the western North Pacific, where the black box denotes the regional forecast domain used in this study. Figure~4b presents the mean track error as a function of forecast lead time for different models. CTL exhibits the largest track errors, especially after 48~hours, when the errors become significantly larger than those of the other models. In contrast, the track errors of HITS, HITS-LPIPS, and ISTM are generally comparable to those of AIFS. At the 120-hour lead time, the average track error is approximately 300~km, indicating a high level of track forecast skill.

Figure~4c compares the mean intensity errors among the models. AIFS consistently exhibits a noticeable negative intensity bias across all lead times. CTL also shows relatively large errors at medium and longer lead times. ISTM improves upon CTL but still systematically underestimates typhoon intensity. In comparison, HITS significantly reduces the overall intensity errors relative to these models. HITS-LPIPS further decreases the intensity errors, with particularly notable improvements during the 36--72~hour forecast period. At the 72-hour lead time, the intensity error of HITS-LPIPS is reduced by 47.8\% compared with AIFS. These batch results further highlight the importance of large-scale dynamical constraints, high-resolution initial conditions, and multi-scale loss functions for improving typhoon structure and intensity forecasts.

To further assess the physical consistency of the simulated typhoon structure, the relationship between maximum wind speed and minimum central pressure ($V_{\max}$--$P_{\min}$) is examined. This wind--pressure relationship is widely used in tropical cyclone studies as a diagnostic indicator of typhoon dynamical balance and intensity structure. Because observational datasets often exhibit scatter between wind and pressure, an empirical fitting curve helps quantify the typical nonlinear relationship between these two quantities and provides a useful benchmark for evaluating model realism. Figure~5a shows the CMA best-track observations, where the red curve represents the empirically fitted wind--pressure relationship
\begin{equation}
\hspace{5cm} V_{\max}=1.5466(1010-P_{\min})^{0.7827}+7.2716
\end{equation}
This empirical relationship reflects the nonlinear strengthening of wind speed as central pressure decreases, particularly during the intense typhoon stage.

The model results reveal clear differences in physical consistency. AIFS (Fig.~5b) shows a noticeable negative bias, indicating systematic underestimation of maximum wind speed. ISTM (Fig.~5c) and CTL (Fig.~5d) also exhibit negative biases. HITS (Fig.~5e) significantly reduces this bias to 0.14~m~s$^{-1}$, producing results that are nearly consistent with the empirical curve. HITS-LPIPS (Fig.~5f) further improves the agreement, with a bias of only 0.04~m~s$^{-1}$, and its scatter distribution aligns most closely with the empirical relationship. These results indicate that the introduction of large-scale dynamical constraints and structure-aware loss function not only improves track prediction but also helps recover the dynamical structure of the typhoon inner core, resulting in a wind--pressure relationship that is more physically consistent.

However, it is also evident that all models show limited skill in predicting very intense typhoons with wind speeds exceeding 50~m~s$^{-1}$ or central pressures lower than 940~hPa. This limitation is likely related to the relatively coarse resolution (9~km) of the HiRes reanalysis used for training, which still tends to underestimate the intensity of the strongest typhoons.
\begin{figure}[!htbp]
\centering
\includegraphics[width=\linewidth]{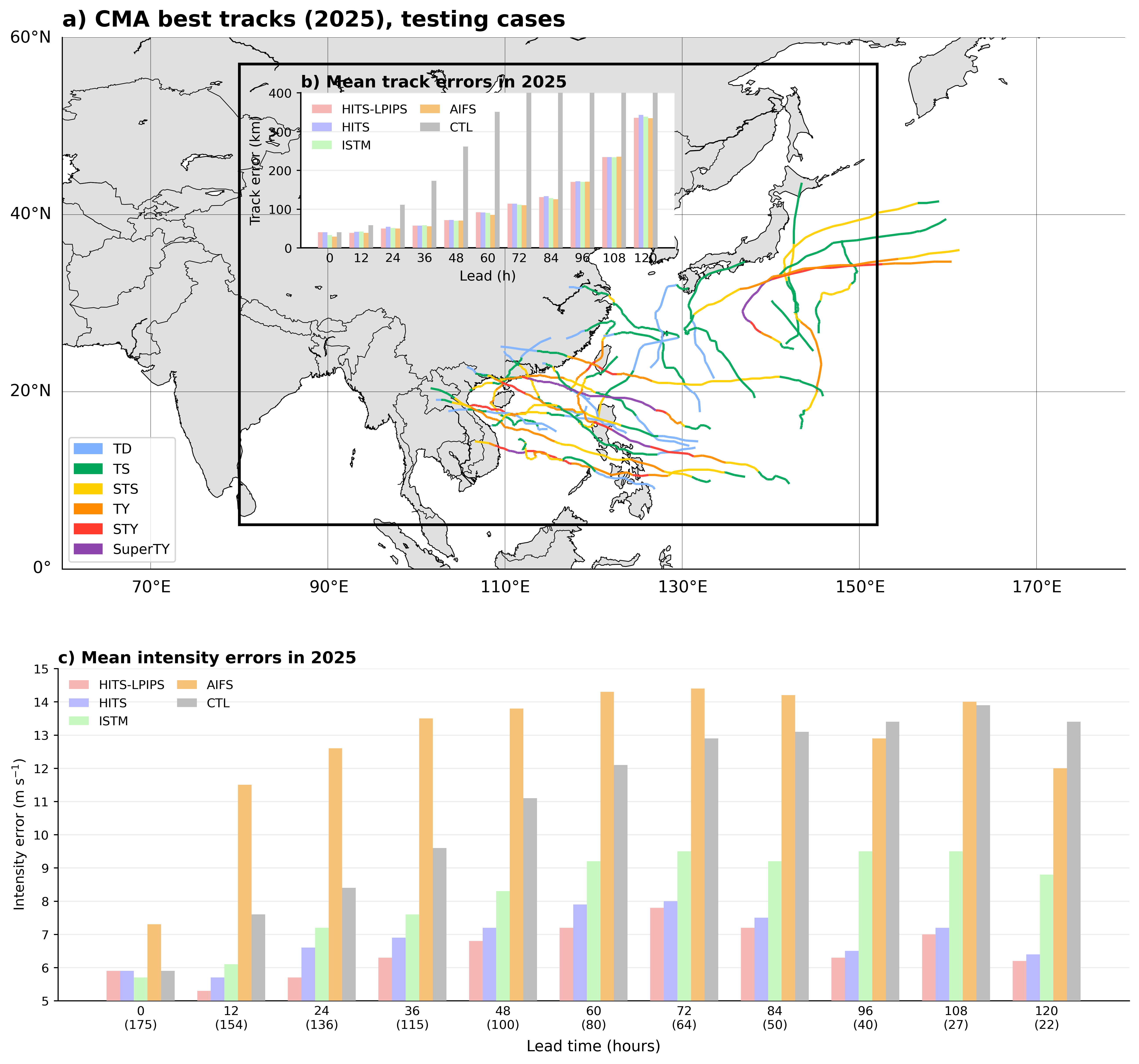}
\caption{\textbf{Spatial distribution of typhoon tracks and forecast errors in 2025.} 
a, CMA best tracks for all typhoon testing cases in 2025 over the western North Pacific, with track segments colored by intensity category (TD, TS, STS, TY, STY and SuperTY). The black rectangle denotes the regional model domain used in this study. The inset shows the mean track errors as a function of forecast lead time for different experiments. 
b, Mean track errors (km) as a function of lead time from 0 to 120~h, comparing HITS-LPIPS, HITS, ISTM, AIFS and CTL. 
c, Mean intensity errors (m~s$^{-1}$) over the same forecast range. Numbers in parentheses along the x-axis indicate the sample size at each lead time. Lower values indicate better forecast performance.}
\label{fig:figure4}
\end{figure}

\begin{figure}[!htbp]
\centering
\includegraphics[width=\linewidth]{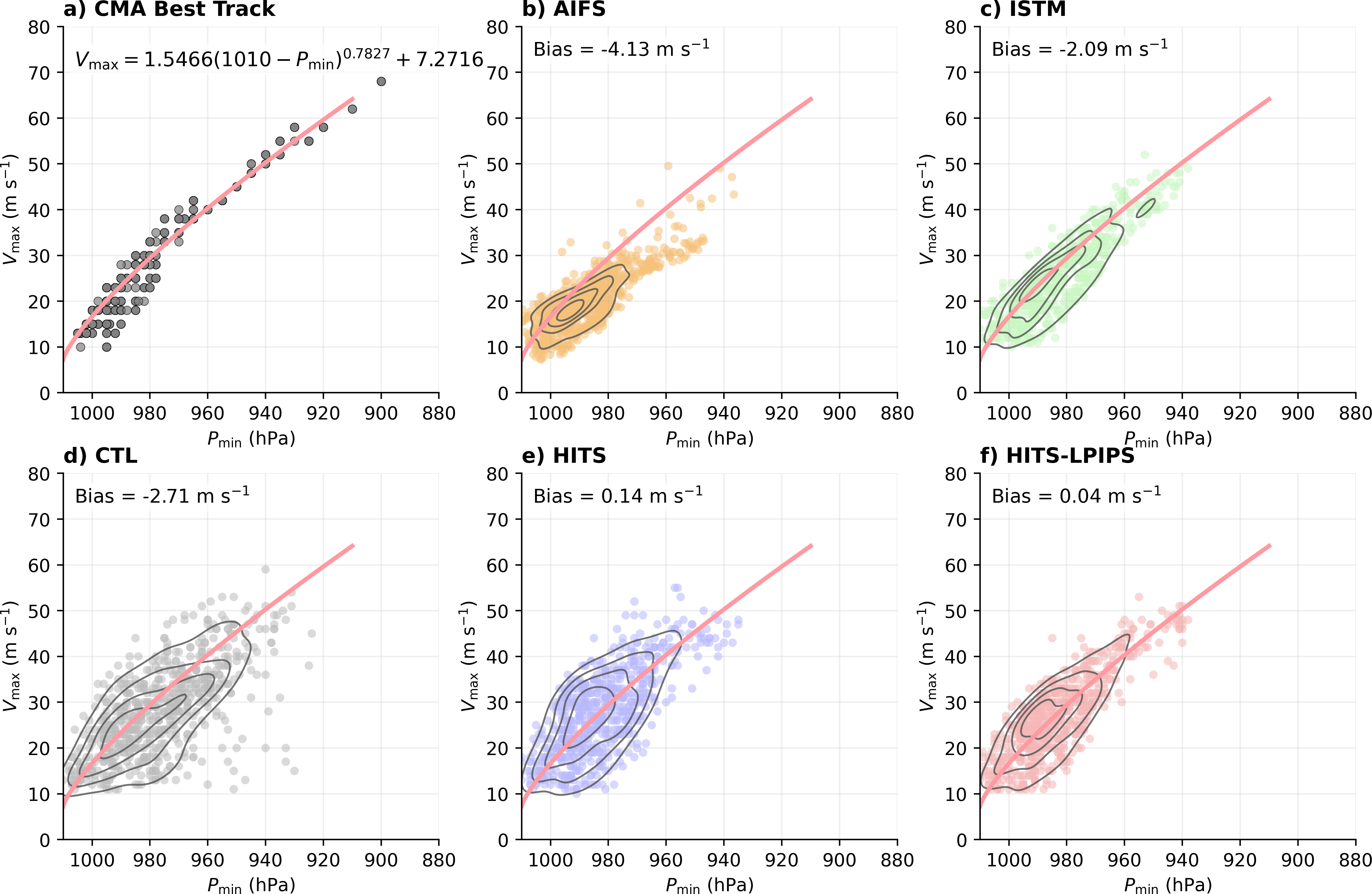}
\caption{\textbf{Relationship between minimum sea-level pressure and maximum wind speed in observations and different forecast experiments.} 
Scatter plots show the relationship between minimum sea-level pressure ($P_{\min}$) and maximum surface wind speed ($V_{\max}$) for a, CMA best-track observations, and forecasts from b, AIFS, c, ISTM, d, CTL, e, HITS, and f, HITS-LPIPS for all typhoons in 2025. The red curve in all panels denotes the empirical pressure--wind relationship fitted from the CMA best-track data. Bias values (m~s$^{-1}$), defined as the mean difference between modelled $V_{\max}$ and the fitted curve, are annotated in each forecast panel.}
\label{fig:figure5}
\end{figure}

\section*{Discussion}
This study demonstrates that dynamically constrained regional AIWP models provide an effective pathway for improving typhoon forecasts at regional scales. By integrating high-resolution regional information with large-scale dynamical guidance from a global AIWP model, the HITS framework successfully bridges the gap between global AI weather prediction and regional mesoscale forecasting. Compared with purely autoregressive regional AIWP models, the introduction of large-scale constraints effectively suppresses medium-range error growth and maintains consistency with the evolving synoptic-scale circulation.

The results also highlight the importance of high-resolution training datasets for extreme weather prediction. Most existing AIWP models are trained on global reanalysis datasets such as ERA5, which cannot adequately represent mesoscale structures critical for typhoon development. The HiRes dataset used in this study provides a higher-resolution representation of typhoon structure and intensity, enabling the regional AI model to learn physically meaningful mesoscale features.

Another key finding is the importance of structure-aware training objectives for representing convective-scale processes. The introduction of the LPIPS perceptual constraint significantly improves the realism of convective structures and precipitation patterns while maintaining quantitative forecast skill. These results suggest that multi-scale feature-based training objectives may provide an effective alternative to generative diffusion approaches for preserving physically coherent structures in atmospheric prediction.

Despite these advances, several limitations remain. In particular, all experiments still underestimate the intensity of the strongest typhoons, which is likely related to the 9-km resolution of the HiRes training dataset. Future work will focus on developing higher-resolution inner-core reanalysis datasets and exploring probabilistic or ensemble extensions of the model to better represent extreme events.

More broadly, this study suggests that hybrid AI forecasting systems combining large-scale dynamical constraints with high-resolution regional learning may represent a promising paradigm for next-generation weather prediction systems.

\section*{Methods}
\subsection*{Architecture of HITS Model}
The training architecture and operational autoregressive workflow of HITS are illustrated in Fig.~6. As shown in Fig.~6a, HITS adopts a hierarchical encoder--decoder framework built upon a U-Transformer backbone. Let $\mathbf{X}_t \in \mathbb{R}^{C\times H\times W}$ denote the HiRes at time $t$, where $C$ is the number of atmospheric variables and $H, W$ are spatial grid dimensions. Let $\mathbf{G}_{t+6} \in \mathbb{R}^{C_g\times H_g\times W_g}$ represent the large-scale background fields at lead time $t+6$~h, derived from ERA5 during training and from AIFS during operational testing.
HITS employs a two-step input to one-step output forecasting scheme, with each step corresponding to a 6-hour interval. The model learns the mapping
\begin{equation}
\hspace{6.5cm}
\hat{\mathbf{X}}_{t+6}=F_{\theta}\left(\mathbf{X}_{t-6:t},\mathbf{G}_{t+6}\right)
\end{equation}
where $\mathbf{X}_{t-6:t}=\{\mathbf{X}_{t-6},\mathbf{X}_t\}$ denotes the two consecutive regional inputs and $F_{\theta}$ is the neural operator parameterized by weights $\theta$. This formulation explicitly couples regional temporal evolution with externally provided large-scale constraints.
In the preprocessing stage (“P” in Fig.~6a), the input tensor is first padded and partitioned into non-overlapping patches of size $p\times p$. The resulting $N=\frac{HW}{p^2}$ patches are linearly embedded into a latent feature space and fed into the encoder. The encoder (“E”) consists of four hierarchical convolutional stages with progressive downsampling, implemented using ResNet-style 2D blocks. This design compresses the high-resolution regional fields into compact multi-scale latent representations while preserving key mesoscale structures. The encoded features are subsequently processed by a Swin-Transformer bottleneck, where multi-head self-attention captures long-range spatial dependencies:
\begin{equation}
\hspace{5cm}
\mathrm{Attention}(\mathbf{Q},\mathbf{K},\mathbf{V})
=\mathrm{Softmax}\left(\frac{\mathbf{Q}\mathbf{K}^{\top}}{\sqrt{d_k}}\right)\mathbf{V}
\end{equation}
A key component of the architecture is the explicit incorporation of large-scale background fields. The large-scale inputs are processed through a parallel preprocessing (“P”) and encoder (“E”) pathway. Their latent representations are injected into the regional bottleneck via a cross-attention module, where regional features serve as queries and large-scale features act as keys and values. Meanwhile, hierarchical features from the large-scale encoder are aligned to the decoder stages through an Adapter module.
During decoding, the decoder (“D”) progressively upsamples the latent features, while Adapter outputs and skip connections from the encoder are fused with the decoder backbone. This design progressively injects large-scale dynamical guidance while retaining mesoscale structural information. Finally, the decoder output is mapped to physical variables through a linear projection layer (“L”), and reshaped back to the original spatial grid through the unpatchify operation (“U”), producing the reconstructed high-resolution fields.
Figure~6b illustrates the operational autoregressive workflow. During training, the large-scale fields $\mathbf{G}_{t+6}$ are obtained from ERA5, whereas in real-time forecasting they are replaced by AIFS predictions at each lead time. Multi-step forecasts are generated recursively,
\begin{equation}
\hspace{5cm}
\hat{\mathbf{X}}_{t+6n}
=F_{\theta}\left(\mathbf{X}_{t+6(n-2):t+6(n-1)},\mathbf{G}_{t+6n}\right),
\quad n\ge1
\end{equation}
where the model ingests the two most recent regional states at each step. For the first step ($n=1$), the input window consists of SHTM analysis at $t-6$ and $t$; thereafter, model outputs are fed back to form a rolling prediction sequence. Crucially, large-scale background fields are externally supplied at every forecast step, ensuring that regional evolution remains dynamically anchored to the evolving large-scale circulation. HITS integrates the dual advantages of high-resolution initial conditions and large-scale dynamical constraints, thereby achieving optimal forecast performance.
\begin{figure}[!htbp]
\centering
\includegraphics[width=\linewidth]{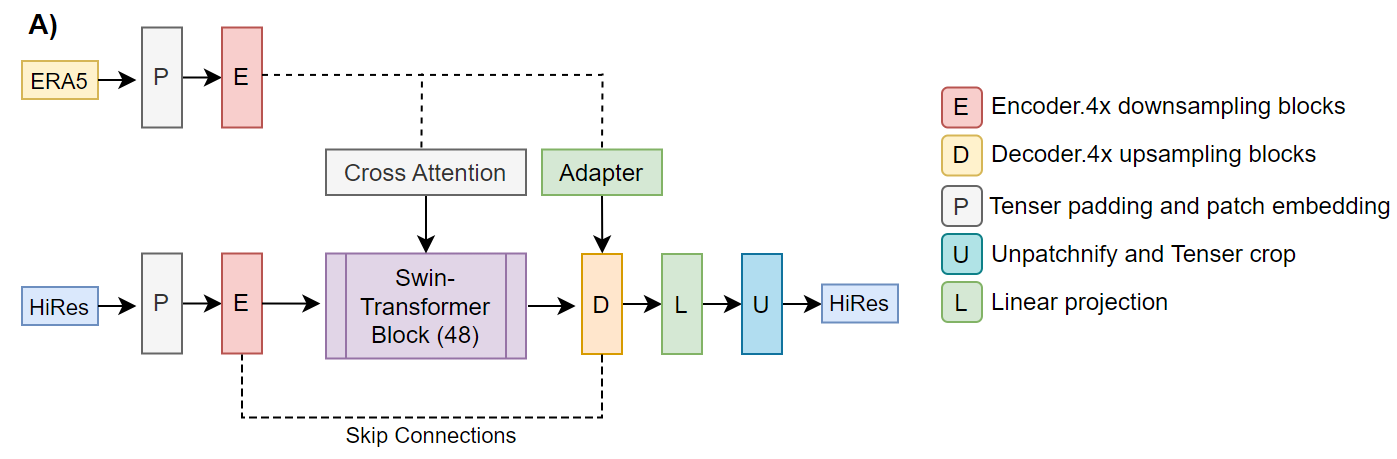}

\vspace{0.2em}

\includegraphics[width=\linewidth]{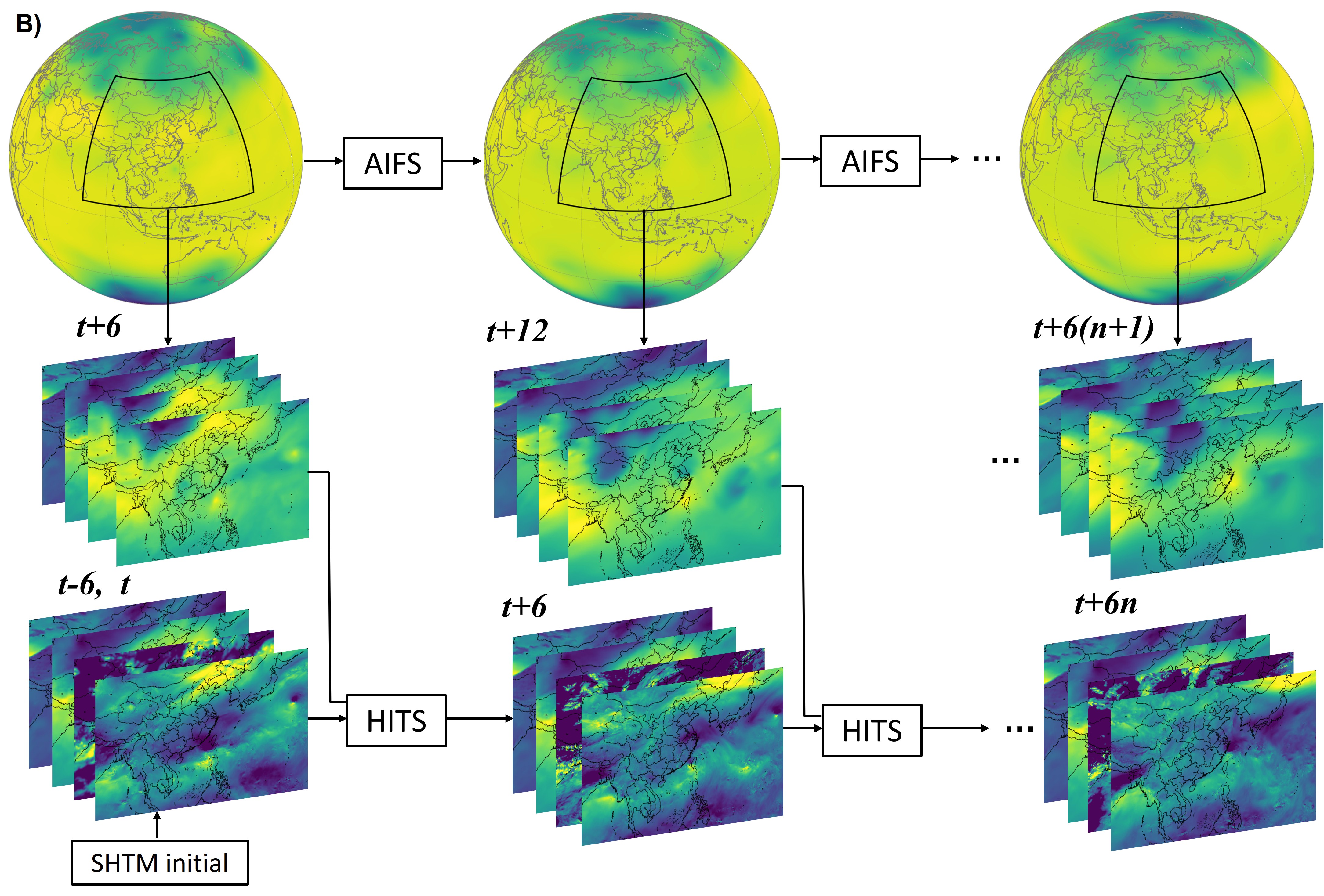}

\caption{\textbf{HITS autoregressive forecasting setup.} 
a, HITS architecture used during training. 
b, Real-time testing workflow of HITS.}
\label{fig:figure6}
\end{figure}

\subsection*{Training configurations}
\subsubsection*{Training dataset}
In this study, the Shanghai Typhoon Model (SHTM) is used to generate the high-resolution typhoon reanalysis dataset (HiRes) for training the regional AI model. SHTM is an operational regional mesoscale prediction system developed by the Shanghai Meteorological Service for tropical cyclone forecasting. It has been running in real time since 2023 as part of the Shanghai Warning and Risk Model System (SWARMS). SHTM is built on the Weather Research and Forecasting (WRF) model and the Gridpoint Statistical Interpolation (GSI) data assimilation system$^{49}$. The model uses a horizontal resolution of 9~km with 56 vertical levels up to 10~hPa. Since 2024, the initial conditions of operational SHTM have been replaced by the ECMWF-HRES analysis, while extensive data assimilation is performed at the forecast initialization time.
The HiRes dataset for typhoons over the Asia--Pacific region during 2018--2025 is generated by continuously integrating SHTM with ERA5 as the background and forcing fields through spectral nudging. Satellite, radar, and TCVitals$^{48-50}$ observations from the China Meteorological Administration are assimilated every six hours (0000, 0600, 1200, and 1800~UTC), producing a high-resolution typhoon reanalysis dataset. The performance of HiRes is evaluated in Fig.~S5 and Fig.~S6. Compared with ERA5, HiRes shows clear improvements in typhoon track, intensity, and wind structure. In particular, the mean intensity error is reduced from 8.97~m~s$^{-1}$ in ERA5 to 4.87~m~s$^{-1}$ in HiRes (Fig.~S6). Further technical details are provided in Niu et~al.~(2025)$^{48}$.
\subsubsection*{Input/Output setups}
As illustrated in Fig.~S7, data from 2018--2023 are used for training and 2024 is used as the validation set for model tuning. The year 2025 is reserved as an independent test period for real-time evaluation because the ECMWF AIFS became operational only in June 2024. Using the 2025 typhoon season therefore enables a consistent evaluation against operational AIFS forecasts. The 2025 testing period includes 27 typhoons spanning multiple intensity categories, from tropical depressions to super typhoons, ensuring a representative assessment of forecast performance. Table~1 summarizes the input/output configuration of HITS during training and testing. HITS adopts a two-step regional input window ($t-6$~h, $t$) and predicts the regional state at $t+6$~h over the domain 80°E--152°E, 5°N--57°N. During training, HiRes fields are used as regional inputs and targets, while ERA5 provides large-scale background fields at $t+6$~h. In operational testing, the regional inputs are replaced by SHTM analysis and the large-scale constraints are provided by AIFS.
Regional variables are defined at 0.1°$\times$0.1° resolution, whereas forcing fields are supplied at 0.25°$\times$0.25° resolution. HITS predicts multi-level atmospheric states at 250, 500, 850, and 925~hPa, including geopotential height (z), temperature (t), horizontal wind components (u, v), and specific humidity (q), together with surface variables and static orography. In addition, the large-scale inputs from AIFS are limited to selected mid-level dynamical variables, specifically z, t, u, and v at 500 and 850~hPa.
\begin{table}[htbp]
\centering
\caption{\textbf{Training and testing configuration of HITS.} A list of the input and output details for HITS.}
\label{tab:table1}

\begin{tabular}{lccc}
\hline
& \multicolumn{2}{c}{Input} & Output \\
\hline

Training I/O & HiRes & ERA5 & HiRes \\
Testing I/O & SHTM analysis & AIFS & HITS \\

Domain & \multicolumn{3}{c}{80°E--152°E, 5°N--57°N} \\

Resolutions & 0.1°$\times$0.1° & 0.25°$\times$0.25° & 0.1°$\times$0.1° \\

\multirow{5}{*}{Pressure level variables}
& z250, z500, z850, z925 & z500, z850 & z250, z500, z850, z925 \\
& t250, t500, t850, t925 & t500, t850 & t250, t500, t850, t925 \\
& u250, u500, u850, u925 & u500, u850 & u250, u500, u850, u925 \\
& v250, v500, v850, v925 & v500, v850 & v250, v500, v850, v925 \\
& qv250, qv500, qv850, qv925 & Total column water vapor & qv250, qv500, qv850, qv925 \\

\multirow{5}{*}{Single level variable}
& \multicolumn{2}{c}{Mean sea level pressure} & Mean sea level pressure \\
& \multicolumn{2}{c}{2m temperature} & 2m temperature \\
& \multicolumn{2}{c}{10m u-component of wind} & 10m u-component of wind \\
& \multicolumn{2}{c}{10m v-component of wind} & 10m v-component of wind \\
& Maximum radar reflectivity & -- & Maximum radar reflectivity \\

Static variable & Orography & -- & -- \\

\hline
\end{tabular}
\end{table}
\subsubsection*{Training-Schedule}
HITS and HITS-LPIPS use almost the same training settings, except for the loss function. We train HITS-LPIPS using distributed data parallelism on eight NVIDIA A100 GPUs (80~GB). Training runs for 100 epochs and takes 86~h in total. Unless otherwise stated, we use a per-GPU batch size of 1 and optimize the model with the Muon optimizer. The optimizer uses a base learning rate of $2\times10^{-4}$, weight decay of $1\times10^{-3}$, momentum of 0.95, and AdamW-style $\beta$ coefficients of $(0.9,0.95)$ with $\epsilon=10^{-8}$. Gradients are clipped with a max-norm of 1.0. Mixed-precision training is supported and enabled when specified.
We adopt a two-stage learning-rate schedule consisting of a linear warm-up followed by cosine annealing. Specifically, the learning rate is linearly warmed up for the first 5 epochs (from $0.1\times$ to $1.0\times$ of the base learning rate), and then annealed using a cosine schedule with a minimum learning rate of $1\times10^{-6}$ for the remaining epochs. Model checkpoints are saved every epoch, and the best-performing checkpoint is selected based on the validation loss. The training objective combines a pixel-wise reconstruction loss and a structure-aware perceptual loss.
We use the mean squared error as the pointwise reconstruction term and add a structure-aware perceptual constraint computed from a pre-trained VGG19 feature extractor. We train and validate the model using distributed samplers to ensure consistent shuffling across GPUs.
\subsubsection*{Muon Optimizer}
To improve training efficiency and stability, we adopt the Muon optimizer instead of AdamW. Muon is a momentum-based adaptive optimizer designed for large-parameter neural networks and has shown faster convergence in high-dimensional settings. Compared with AdamW, Muon relies primarily on momentum with adaptive rescaling, reducing dependence on second-moment normalization while maintaining stable update magnitudes. As shown in Fig.~S8, Muon reaches the target training loss of 0.05 about twice as fast as AdamW ($\approx2.0\times$ speedup) and converges to a lower final loss, indicating improved optimization efficiency for the HITS model.
\subsubsection*{Learned Perceptual Image Patch Similarity (LPIPS)}
To better represent mesoscale and convective-scale structures in typhoon forecasts, we incorporate LPIPS into the training objective. Unlike pixel-wise losses such as L1 or L2, which penalize pointwise differences in physical space, LPIPS evaluates similarity in a learned feature space extracted by a fixed pre-trained convolutional neural network. Given a prediction field $\mathbf{x}$ and a reference field $\mathbf{y}$, both are passed through the pretrained feature extractor to obtain multi-layer feature representations corresponding to different spatial scales. Differences between the two fields are then computed in feature space rather than directly at the pixel level. The LPIPS loss is defined as
\begin{equation}
\hspace{5cm}
L_{\mathrm{LPIPS}}(\mathbf{x},\mathbf{y})
=\sum_{l} w_l
\left\|
\hat{f}_l(\mathbf{x})-\hat{f}_l(\mathbf{y})
\right\|_2^2
\end{equation}
where $\hat{f}_l(\cdot)$ denotes the normalized feature map at layer $l$, and $w_l$ are learned layer-wise weights. As illustrated in Fig.~S9, multi-scale feature differences are aggregated to form a perceptual similarity metric. This design enables the model to penalize structural discrepancies in organized convective bands, eyewall structures, and mesoscale precipitation patterns, rather than focusing solely on pixel-level amplitude errors.
The total training loss of HITS-LPIPS is therefore defined as
\begin{equation}
\hspace{6.5cm}
\mathcal{L}=L_{\mathrm{rec}}+\lambda L_{\mathrm{LPIPS}}
\end{equation}
where $L_{\mathrm{rec}}=\|\mathbf{x}-\mathbf{y}\|_1/(CHW)$ denotes the mean absolute reconstruction loss over all channels and spatial grids, and $\lambda=0.1$ balances structural fidelity and quantitative accuracy. This composite objective reduces excessive smoothing while maintaining large-scale physical consistency.
\subsection*{Evaluation Metrics}
\subsubsection*{Root Mean Square Error.}
For gridded atmospheric variables (e.g., temperature, wind, and specific humidity), forecast accuracy is evaluated using the root mean square error (RMSE). Given forecast field $F(x,y,t)$ and reference field $O(x,y,t)$ at lead time $t$, the RMSE is defined as
\begin{equation}
\hspace{5cm}
\mathrm{RMSE}(t)
=
\sqrt{
\frac{1}{M}
\sum_{j=1}^{M}
\left(F_j(t)-O_j(t)\right)^2
}
\end{equation}
where $M$ is the total number of grid points in the verification domain. RMSE provides an aggregate measure of amplitude errors and is sensitive to both systematic bias and random fluctuations.
\subsubsection*{Fractions Skill Score.}
To evaluate spatial precipitation and reflectivity structures, we employ the Fractions Skill Score (FSS), which measures neighborhood-scale agreement between forecast and observation. For a given threshold $T$ and neighborhood size $L$, the fractional coverage within a local window $\mathcal{N}_{x,y}$ is computed as
\begin{equation}
\hspace{5cm}
\begin{aligned}
P_f(x,y)&=\frac{1}{L^2}\sum_{(i,j)\in\mathcal{N}_{x,y}}\mathbf{1}\big(F(i,j)\ge T\big),\\
P_o(x,y)&=\frac{1}{L^2}\sum_{(i,j)\in\mathcal{N}_{x,y}}\mathbf{1}\big(O(i,j)\ge T\big).
\end{aligned}
\end{equation}
where $F$ and $O$ denote forecast and observed fields, respectively, and $\mathbf{1}(\cdot)$ is the indicator function. In this study, the neighborhood window size is set to $L=9$, corresponding to a $9\times9$ grid-point window.The FSS is then defined as
\begin{equation}
\hspace{5cm}
\mathrm{FSS}
=
1-
\frac{
\sum_{x,y}\left(P_f(x,y)-P_o(x,y)\right)^2
}{
\sum_{x,y}P_f(x,y)^2+\sum_{x,y}P_o(x,y)^2
}
\end{equation}
FSS ranges from 0 to 1, with 1 indicating perfect spatial agreement. By varying $L$ and $T$, FSS quantifies forecast skill across spatial scales, particularly for convective precipitation and radar reflectivity structures.
\subsubsection*{Mean Typhoon Track and Intensity Error.}
To evaluate tropical cyclone forecast performance, we compute the mean track error and mean intensity error relative to CMA best-track. The mean track error at lead time $t$ is defined as
\begin{equation}
\hspace{6.5cm}
E_{\mathrm{track}}(t)
=
\frac{1}{N}
\sum_{i=1}^{N}
D_i(t)
\end{equation}
where $D_i(t)$ denotes the horizontal position error (in km) between the forecast and observed typhoon center for case $i$, and $N$ is the total number of samples.
The mean intensity error at lead time $t$ is defined as
\begin{equation}
\hspace{6cm}
E_{\mathrm{int}}(t)
=
\frac{1}{N}
\sum_{i=1}^{N}
\left|I_{f,i}(t)-I_{o,i}(t)\right|
\end{equation}
where $I$ represents the maximum 10~m wind speed ($V_{\max}$). These metrics quantify forecast skill in predicting tropical cyclone motion and intensity evolution.

\section*{Data availability}

The ERA5 reanalysis data used for training are publicly available from the Copernicus Climate Data Store at \url{https://cds.climate.copernicus.eu/datasets}. The operational AIFS analysis data are available at \url{https://ecmwf-forecasts.s3.amazonaws.com}. The HiRes dataset is available from the authors upon reasonable request.

\section*{Code availability}

The HITS training codes are available at \url{https://github.com/ZeyiNiu/HITS}.

\section*{Acknowledgements}

This research was supported by the National Youth Science Foundation of China (Grant No.~4240050560), the Special Project--Original Exploration (Grant No.~42450163), and the Typhoon Scientific and Technological Innovation Group of the China Meteorological Administration (CMA2023ZD06).
\section*{Author contributions}
The authors contributed equally to all aspects of the article.

\section*{References}

\noindent\textbf{1.} Bauer, P., Thorpe, A. \& Brunet, G. The quiet revolution of numerical weather prediction. \textit{Nature} \textbf{525}, 47--55 (2015).

\noindent\textbf{2.} Brotzge, J. A., Berchoff, D., Carlis, D. L., Carr, F. H., Carr, R. H., Gerth, J. J. \textit{et al.} Challenges and opportunities in numerical weather prediction. \textit{Bull. Am. Meteorol. Soc.} \textbf{104}, E698--E705 (2023).

\noindent\textbf{3.} Fischer, E. M. \& Knutti, R. Observed heavy precipitation increase confirms theory and early models. \textit{Nat. Clim. Change} \textbf{6}, 986--991 (2016).

\noindent\textbf{4.} Yu, H., Chen, G., Wong, W. K., Vigh, J. L., Pan, C. K., Lu, X. \textit{et al.} WMO typhoon landfall forecast demonstration project (2010–22): A decade of transition from track forecasts to impact forecasts. \textit{Bull. Am. Meteorol. Soc.} \textbf{105}, E1320--E1349 (2024).

\noindent\textbf{5.} Dee, D. P., Uppala, S., Simmons, A. J., Berrisford, P., Poli, P., Kobayashi, S. \textit{et al.} The ERA-Interim reanalysis: Configuration and performance of the data assimilation system. \textit{Q. J. R. Meteorol. Soc.} \textbf{137}, 553--597 (2011).

\noindent\textbf{6.} Hersbach, H., Bell, B., Berrisford, P., Hirahara, S., Horányi, A., Muñoz-Sabater, J. \textit{et al.} The ERA5 global reanalysis. \textit{Q. J. R. Meteorol. Soc.} \textbf{146}, 1999--2049 (2020).

\noindent\textbf{7.} Kalnay, E. Atmospheric Modeling, Data Assimilation and Predictability. Cambridge Univ. Press (2003).

\noindent\textbf{8.} Wedi, N. P., Polichtchouk, I., Dueben, P., Anantharaj, V. G., Bauer, P., Boussetta, S. \textit{et al.} A baseline for global weather and climate simulations at 1 km resolution. \textit{J. Adv. Model. Earth Syst.} \textbf{12}, e2020MS002192 (2020).

\noindent\textbf{9.} Stensrud, D. J. Parameterization Schemes: Keys to Understanding Numerical Weather Prediction Models. Cambridge Univ. Press (2007).

\noindent\textbf{10.} Hong, S. Y. \& Dudhia, J. Next-generation numerical weather prediction: Bridging parameterization, explicit clouds, and large eddies. \textit{Bull. Am. Meteorol. Soc.} \textbf{93}, ES6--ES9 (2012).

\noindent\textbf{11.} Geer, A. J. \& Bauer, P. Observation errors in all-sky data assimilation. \textit{Q. J. R. Meteorol. Soc.} \textbf{137}, 2024--2037 (2011).

\noindent\textbf{12.} Geer, A. J., Lonitz, K., Weston, P., Kazumori, M., Okamoto, K., Zhu, Y. \textit{et al.} All-sky satellite data assimilation at operational weather forecasting centres. \textit{Q. J. R. Meteorol. Soc.} \textbf{144}, 1191--1217 (2018).

\noindent\textbf{13.} Geer, A. J. Learning earth system models from observations: Machine learning or data assimilation? \textit{Philos. Trans. R. Soc. A} \textbf{379}, 20200089 (2021).

\noindent\textbf{14.} Eyre, J. R., Bell, W., Cotton, J., English, S. J., Forsythe, M., Healy, S. B. \textit{et al.} Assimilation of satellite data in numerical weather prediction. Part II: Recent years. \textit{Q. J. R. Meteorol. Soc.} \textbf{148}, 521--556 (2022).

\noindent\textbf{15.} Selz, T., Riemer, M. \& Craig, G. C. The transition from practical to intrinsic predictability of midlatitude weather. \textit{J. Atmos. Sci.} \textbf{79}, 2013--2030 (2022).

\noindent\textbf{16.} Zhang, F., Sun, Y. Q., Magnusson, L., Buizza, R., Lin, S. J., Chen, J. H. \textit{et al.} What is the predictability limit of midlatitude weather? \textit{J. Atmos. Sci.} \textbf{76}, 1077--1091 (2019).

\noindent\textbf{17.} Ben Bouallègue, Z., Clare, M. C., Magnusson, L., Gascon, E., Maier-Gerber, M., Janoušek, M. \textit{et al.} The rise of data-driven weather forecasting. \textit{Bull. Am. Meteorol. Soc.} \textbf{105}, E864--E883 (2024).

\noindent\textbf{18.} Dueben, P. D. \& Bauer, P. Challenges and design choices for global weather models based on machine learning. \textit{Geosci. Model Dev.} \textbf{11}, 3999--4009 (2018).

\noindent\textbf{19.} Scher, S. Toward data-driven weather forecasting. \textit{Geophys. Res. Lett.} \textbf{45}, 12616--12622 (2018).

\noindent\textbf{20.} Weyn, J. A., Durran, D. R. \& Caruana, R. Can machines learn to predict weather? \textit{J. Adv. Model. Earth Syst.} \textbf{11}, 2680--2693 (2019).

\noindent\textbf{21.} Rasp, S., Dueben, P. D., Scher, S., Weyn, J. A., Mouatadid, S. \& Thuerey, N. WeatherBench. \textit{J. Adv. Model. Earth Syst.} \textbf{12}, e2020MS002203 (2020).

\noindent\textbf{22.} Bi, K., Xie, L., Zhang, H., Chen, X., Gu, X. \& Tian, Q. Accurate medium-range global weather forecasting with 3D neural networks. \textit{Nature} \textbf{619}, 533--538 (2023).

\noindent\textbf{23.} Pathak, J., Subramanian, S., Harrington, P., Raja, S., Chattopadhyay, A., Mardani, M. \textit{et al.} FourCastNet. arXiv:2202.11214 (2022).

\noindent\textbf{24.} Lam, R., Sanchez-Gonzalez, A., Willson, M., Wirnsberger, P., Fortunato, M., Alet, F. \textit{et al.} Learning skillful medium-range global weather forecasting. \textit{Science} \textbf{382}, 1416--1421 (2023).

\noindent\textbf{25.} Chen, K., Han, T., Ling, F., Gong, J., Bai, L., Wang, X. \textit{et al.} Extending operational weather forecasting beyond 10 days. \textit{Commun. Earth Environ.} \textbf{6}, 518 (2025).

\noindent\textbf{26.} Chen, L., Zhong, X., Zhang, F., Qian, W., Liu, J. \& Li, H. FuXi: A cascade ML forecasting system. \textit{npj Clim. Atmos. Sci.} \textbf{6}, 190 (2023).

\noindent\textbf{27.} Lang, S., Alexe, M., Chantry, M., Dramsch, J., Pinault, F., Raoult, B. \textit{et al.} AIFS–ECMWF's data-driven forecasting system. arXiv:2406.01465 (2024).

\noindent\textbf{28.} Kochkov, D., Yuval, J., Langmore, I., Norgaard, P., Smith, J., Mooers, G. \textit{et al.} Neural general circulation models for weather and climate. \textit{Nature} \textbf{632}, 1060--1066 (2024).

\noindent\textbf{29.} Price, I., Sanchez-Gonzalez, A., Alet, F., Andersson, T. R., El-Kadi, A., Masters, D. \textit{et al.} Probabilistic weather forecasting with machine learning. \textit{Nature} \textbf{637}, 84--90 (2025).

\noindent\textbf{30.} Bodnar, C., Bruinsma, W. P., Lucic, A., Stanley, M., Allen, A., Brandstetter, J. \textit{et al.} A foundation model for the Earth system. \textit{Nature} \textbf{641}, 1180--1187 (2025).

\noindent\textbf{31.} Allen, A., Markou, S., Tebbutt, W., Requeima, J., Bruinsma, W. P., Andersson, T. R. \textit{et al.} End-to-end data-driven weather prediction. \textit{Nature} (2025).

\noindent\textbf{32.} Olivetti, L. \& Messori, G. Do data-driven models beat numerical models in forecasting extremes? EGUsphere (2024).

\noindent\textbf{33.} Sun, Y. Q., Hassanzadeh, P., Zand, M., Chattopadhyay, A., Weare, J. \& Abbot, D. S. Can AI weather models predict gray swan tropical cyclones? \textit{Proc. Natl Acad. Sci. USA} \textbf{122}, e2420914122 (2025).

\noindent\textbf{34.} Xu, H., Zhao, Y., Zhao, D., Duan, Y. \& Xu, X. Exploring typhoon intensity forecasting by integrating AI with NWP. \textit{npj Clim. Atmos. Sci.} \textbf{8}, 38 (2025).

\noindent\textbf{35.} Liu, H. Y., Tan, Z. M., Wang, Y., Tang, J., Satoh, M., Lei, L. \textit{et al.} A hybrid ML/physics framework for tropical cyclones. \textit{JGR Mach. Learn. Comput.} \textbf{1}, e2024JH000207 (2024).

\noindent\textbf{36.} Niu, Z., Huang, W., Zhang, L., Deng, L., Wang, H., Yang, Y. \textit{et al.} Improving typhoon predictions using spectral nudging and ML. \textit{Earth Space Sci.} \textbf{12}, e2024EA003952 (2025).

\noindent\textbf{37.} Husain, S. Z. \textit{et al.} Leveraging data-driven models for improving NWP skill. \textit{Wea. Forecasting} \textbf{40}, 1749--1771 (2025).

\noindent\textbf{38.} Lai, S. K., He, Y., Chan, P. W., Kerns, B. W., Chen, S. S. \& Su, H. Towards skillful tropical cyclone forecasting by AI-driven models. \textit{Meteorol. Appl.} \textbf{32}, e70109 (2025).

\noindent\textbf{39.} Mardani, M., Brenowitz, N., Cohen, Y. \textit{et al.} Residual corrective diffusion modeling for km-scale atmospheric downscaling. \textit{Commun. Earth Environ.} \textbf{6}, 124 (2025).

\noindent\textbf{40.} Zhang, E., Su, H., Chan, P. W., Zhai, C., Zhou, W., Xu, W. \textit{et al.} Deep learning-based multisource satellite data fusion. \textit{JGR Mach. Learn. Comput.} \textbf{2}, e2025JH000792 (2025).

\noindent\textbf{41.} Adamov, S., Oskarsson, J., Denby, L., Landelius, T., Hintz, K., Christiansen, S. \textit{et al.} Building machine learning limited area models. arXiv:2504.09340 (2025).

\noindent\textbf{42.} Oskarsson, J., Landelius, T. \& Lindsten, F. Graph-based neural weather prediction. arXiv:2309.17370 (2023).

\noindent\textbf{43.} Oskarsson, J., Landelius, T., Deisenroth, M. \& Lindsten, F. Probabilistic forecasting with hierarchical GNN. \textit{Adv. Neural Inf. Process. Syst.} \textbf{37}, 41577--41648 (2024).

\noindent\textbf{44.} Nipen, T. N., Haugen, H. H., Ingstad, M. S., Nordhagen, E. M., Salihi, A. F. S., Tedesco, P. \textit{et al.} Regional data-driven weather modeling. \textit{Artif. Intell. Earth Syst.} (2025).

\noindent\textbf{45.} Xu, P., Zheng, X., Gao, T., Wang, Y., Yin, J., Zhang, J. \textit{et al.} An AI-based limited area weather model. \textit{Commun. Earth Environ.} \textbf{6}, 372 (2025).

\noindent\textbf{46.} Pathak, J., Cohen, Y., Garg, P., Harrington, P., Brenowitz, N., Durran, D. \textit{et al.} Kilometer-scale convection-allowing model emulation. \textit{Sci. Adv.} \textbf{12}, eadv0423 (2026).

\noindent\textbf{47.} Abdi, D., Jankov, I., Madden, P., Vargas, V., Smith, T. A., Frolov, S. \textit{et al.} HRRRCast: A data-driven emulator for regional forecasting. \textit{Artif. Intell. Earth Syst.} (2026).

\noindent\textbf{48.} Niu, Z., Huang, W., Huang, S., Qin, B., Yang, M., Sun, H. \textit{et al.} Intelligent Shanghai Typhoon Model (ISTM). arXiv:2508.16851 (2025).

\noindent\textbf{49.} Niu, Z., Huang, W., Yang, Y., Yang, M., Deng, L., Wang, H. \textit{et al.} Evaluating the Shanghai Typhoon Model against ML weather models.

\noindent\textbf{50.} Niu, Z., Zhang, L., Yang, Y., Han, Y., Li, H., Wang, D. \textit{et al.} Assimilating FY-4B AGRI water vapor channels in SHTM. \textit{IEEE J. Sel. Top. Appl. Earth Obs. Remote Sens.} \textbf{18}, 3599--3610 (2024).

\noindent\textbf{51.} Huo, Z., Liu, Y., Taylor, J., Zhou, Y., Amemiya, A., Fan, H. \& Miyoshi, T. Incremental analysis updates in convective-scale EnKF. \textit{J. Adv. Model. Earth Syst.} \textbf{17}, e2024MS004802 (2025).

\noindent\textbf{52.} Holt, C., Szunyogh, I., Gyarmati, G., Leidner, S. M. \& Hoffman, R. N. Assimilation of tropical cyclone observations. \textit{Mon. Wea. Rev.} \textbf{143}, 3956--3980 (2015).

\section*{Supplementary information}
Figs. S1 to S9

\end{document}